\documentclass[conference]{IEEEtran}

\usepackage[utf8]{inputenc}
\usepackage[T1]{fontenc}
\usepackage{microtype}

\usepackage{amsmath}
\usepackage{graphicx}
\usepackage{float}
\usepackage{xcolor}
\usepackage{xspace}
\usepackage{ifthen}
\usepackage{soul}
\usepackage{comment}
\usepackage{booktabs} 
\usepackage[plain]{algorithm}
\usepackage{algpseudocode}
\usepackage[font=footnotesize]{caption}
\usepackage{capt-of}
\usepackage{hyperref}
\usepackage{listings}
\usepackage{flushend}

\newcommand{\system}[0]{UNIT\xspace}
\newcommand{\rewriter}[0]{Rewriter\xspace}

\newcommand{\showcomments}{yes}
\newcommand\ignore[1]{
    \ifthenelse{\equal{\showcomments}{yes}}{{}}{\ignorespaces}
}

\definecolor{mygreen}{rgb}{0,0.6,0}
\definecolor{mygray}{rgb}{0.5,0.5,0.5}
\definecolor{mymauve}{rgb}{0.58,0,0.82}

\begin{document}

\title{\system: Unifying Tensorized Instruction Compilation}

\date{}
\author{
\IEEEauthorblockN{Jian Weng\IEEEauthorrefmark{1}\IEEEauthorrefmark{2},
Animesh Jain\IEEEauthorrefmark{2},
Jie Wang\IEEEauthorrefmark{1}\IEEEauthorrefmark{2},
Leyuan Wang\IEEEauthorrefmark{2},
Yida Wang\IEEEauthorrefmark{2},
Tony Nowatzki\IEEEauthorrefmark{1}
}
\vspace{0.05in}
\IEEEauthorblockA{\IEEEauthorrefmark{1}University of California,
Los Angeles, USA  ~~~~  \IEEEauthorrefmark{2}Amazon Web Services, USA}
\vspace{0.05in}
\IEEEauthorblockA{
\texttt{\{jian.weng,jiewang,tjn\}@cs.ucla.edu}~~~~\texttt{\{janimesh,wangleyu,wangyida\}@amazon.com}
}
\vspace{-0.2in}

}

\maketitle

\begingroup\renewcommand\thefootnote{\IEEEauthorrefmark{1}\IEEEauthorrefmark{2}}
\footnotetext{Work done during Jian and Jie's internship at AWS.}
\endgroup

\thispagestyle{empty}
\pagestyle{empty}
\setlength{\textfloatsep}{0pt}

\begin{abstract}

Because of the increasing demand for intensive computation in deep neural networks, researchers have developed both hardware and software mechanisms to reduce the compute and memory burden. A widely adopted approach is to use mixed precision data types. However, it is hard to benefit from mixed precision without hardware specialization because of the overhead of data casting. Recently, hardware vendors offer \emph{tensorized} instructions specialized for mixed-precision tensor operations, such as Intel VNNI, Nvidia Tensor Core, and ARM DOT. These instructions involve a new computing idiom, which reduces multiple low precision elements into one high precision element. The lack of compilation techniques for this emerging idiom makes it hard to utilize these instructions. In practice, one approach is to use vendor-provided libraries for computationally-intensive kernels, but this is inflexible and prevents further optimizations. Another approach is to manually write hardware intrinsics, which is error-prone and difficult for programmers. Some prior works tried to address this problem by creating compilers for each instruction. This requires excessive efforts when it comes to many tensorized instructions.

In this work, we develop a compiler framework, \system, to \underline{uni}fy the compilation for \underline{t}ensorized instructions. The key to this approach is a unified semantics abstraction which makes the integration of new instructions easy, and the reuse of the analysis and transformations possible.
Tensorized instructions from different platforms can be compiled via \system with moderate effort for favorable performance. Given a tensorized instruction and a tensor operation, \system automatically detects the applicability of the instruction, transforms the loop organization of the operation, and rewrites the loop body to take advantage of the tensorized instruction. According to our evaluation, \system is able to target various mainstream hardware platforms. The generated end-to-end inference model achieves 1.3$\times$ speedup over Intel oneDNN on an x86 CPU, 1.75$\times$ speedup over Nvidia cuDNN on an Nvidia GPU, and 1.13$\times$ speedup over a carefully tuned TVM solution for ARM DOT on an ARM CPU.

\end{abstract}

\section{Introduction} \label{sec:intro}
Dense tensor operations like matrix multiplication (Matmul) and convolution (Conv) have long been the workhorses in many domains, including deep learning workloads \cite{imagenet}.
The popularity of deep learning means that aggressively optimizing these operations has a high payoff.
Essentially, Matmul and Conv are a series of multiply-accumulate (MAC) operations, which perform accumulation over a number of elementwise multiplications.

To capture the reduction behavior and perform it more efficiently,
recent general-purpose processors offer native tensor operation specialized instructions (hereinafter referred to as \emph{tensorized instructions}), like Intel VNNI~\cite{vnni-page}, Nvidia Tensor Core~\cite{tc-page}, and ARM DOT~\cite{vdot-page}. 
Unlike the conventional SIMD instructions, after performing elementwise arithmetic operations, these instructions introduce a ``horizontal computation'' to accumulate elementwise results.
Further, tensorized instructions are often mixed-precision, meaning that elementwise operations
use less precise and lower bitwidth operands (e.g., \texttt{fp16} and \texttt{int8}), while accumulation
occurs with higher bitwidth, where it is needed.  This offers a good balance between data width and precision that is generally sufficient for deep learning workloads~\cite{mixed_precision, tflite_quant}, and enables the use of quantized data types.

Mixed-precision is difficult to express in a single SIMD
instruction, because the output vector width is different than the input vector width. In most ISAs this paradigm requires multiple SIMD
instructions to express.  In a tensorized instruction, by definition there are fewer outputs, so allocating more bitwidth to them for the output vector is natural.
In addition, tensorized instructions sometimes reuse the same inputs multiple times, which reduces the required register file bandwidth.
Overall, tensorized instructions offer significant advantages over SIMD for executing MACs.

While promising, the absence of appropriate compilation techniques limit c the applicability of these tensorized instructions. Conventional SIMD instructions are vector instructions, so industry standard compilers only try parallelizing the innermost loops. In addition, it is difficult for the high-level language programmer to express the compute flow in a tensorization-friendly way and hint the compiler to try tensorization upon a loop nest, because the dependency of reduction is more complicated and error-prone.

In practice, there are normally two options to leverage tensorized instructions.
One way is to call the vendor-provided libraries such as Intel oneDNN~\cite{intel-onednn}, Nvidia cuBLAS and cuDNN~\cite{nv-cudnn}, which provides highly optimized performance in some pre-defined single kernels using tensorized instructions~\cite{xmlowbit,dem-tc}.
However, it also brings inflexibility when it comes to new workloads or when further performance exploitation is desired.
The other option is to manually write assembly intrinsics, which sets a high bar to ordinary developers and hence lacks productivity.
Some prior works tried to solve this problem by developing a compiler~\cite{nv-auto-tc,polydl} for each instruction. This requires too much effort when there are many tensorized instructions, both within and across hardware platforms.

\noindent \textbf{Our Goal:} Although different processors may provide different tensorized instructions, in the context of deep learning workloads, we observe that these instructions essentially handle a similar compute pattern, i.e., elementwise multiplication and then horizontal accumulation.  They
primarily differ in the number of elementwise computation lanes and the accepting data types.
Therefore, we aim to develop a unified approach to compile these tensorized instructions on multiple platforms to optimize the tensor operations in deep learning workloads.
Our techniques are extensible to the tensorized instructions with other data types and operations as well.

\noindent \textbf{Challenges: } There are several challenges to attain a unified compilation pipeline:
\begin{itemize}
    \item \emph{Instructions Integration:} Instead of building a new specialized compiler for each new instruction, it is desirable to create a unified and extensible compilation flow; 
    \item \emph{Detecting the applicability:} Given a tensorized instruction, a first question is whether and how this instruction can be applied to the target tensor operation, which may require loop reorganization to make it applicable;
    \item \emph{Code rewriting:} When applicable,
    the compiler must determine how the loops involved should be rewritten by the tensorized instruction, and how the loops should be rearranged to achieve high performance.
\end{itemize}

\noindent \textbf{Our Insight:} We envision that the key to addressing these three challenges is to have a unified semantics abstraction for tensorized instructions so that the analysis and transformation can also be unified.

This paper presents \system, an end-to-end compilation pipeline to surmount the above three challenges.
\system takes the tensorized instructions (e.g., Intel VNNI instructions on CPUs, or Nvidia Tensor Core instructions on GPUs) and a deep learning model as input, lowers the tensor operations of the model into loop-based IRs to identify the tensorizable components, and inserts the tensorized instructions by transforming and rewriting the loop.  It achieves high performance for tensor operations, and consequently, model inference.
To the best of our knowledge, this is the first work to tackle tensorized instruction compilation and optimization with a unified solution.
\system not only achieves high performance for single tensor operations, but also provides desirable model inference latency in practice.


\noindent \textbf{Key Results:} According to our evaluation, \system is expressive enough to target many tensorized instructions on multiple hardware platforms, including Intel VNNI, Nvidia Tensor Core, and ARM DOT.
The generated programs for end-to-end model inference are 1.3$\times$ and 1.75$\times$ faster than the solutions backed up by Intel oneDNN and Nvidia cuDNN on CPU and GPU, respectively.
In addition, \system can be extended to new tensorized instructions with moderate effort.
Although we designed \system to target Intel CPUs and Nvidia GPUs, on an ARM Cortex A-72 CPU with DOT instructions, \system achieves up to 1.13$\times$ speedup against a carefully manual tuned solution.

To sum up, our contribution is an end-to-end compilation pipeline of tensorized instructions for deep learning workloads, which includes:
\begin{itemize}
    \item A unified abstraction for tensorized instructions.
    \item An algorithm that detects the applicability of these tensorized instructions.
    \item A rewriting and tuning mechanism that looks for favorable loop transformations of the tensor operations to plug in the tensorized instructions for high performance.
\end{itemize}

\noindent \textbf{Paper Organization:} We first introduce the background and challenges of tensorized compilation in Section~\ref{sec:bg}. The design of \system is presented in Section~\ref{sec:tensorizer}. We explain the implementation details in Section~\ref{sec:impl}. We clarify our experiment methodology in Section~\ref{sec:method}, and evaluate our work in Section~\ref{sec:eval}. Finally, we discuss the related work in Section~\ref{sec:rel}.
\section{Background} \label{sec:bg}

\system is an end-to-end compilation pipeline capable of automatically mapping tensorized instructions to the deep learning tensor operations. It defines the tensorized instruction's semantics using a suitable intermediate representation (IR) and inserts them in proper places of the program of tensor operations. In this section, we give an overview of popular mixed precision tensorized instructions, followed by the limitations of existing solutions in automatic mapping of these tensorized instructions. Finally, we discuss the background of tensor domain specific language and the multi-level intermediate representation.

\subsection{Mixed Precision Tensorized Instructions}
Deep learning is computationally expensive, requiring substantial compute and memory resources. As deep learning becomes more pervasive, researchers are designing both software and hardware techniques to reduce the compute and memory burden. A widely adopted approach in this context is using mixed precision for expensive operations, e.g., convolution or dense operations~\cite{mixed_precision, tflite_quant}. In practice, this means representing 32-bit floating point (\texttt{fp32}) operands with a lower bitwidth datatype - 16-bit floating point numbers (\texttt{fp16}) or 8/16-bit integer numbers (\texttt{int8}, \texttt{int16}). To keep the accuracy in check, it is helpful to accumulate the results in higher precision (\texttt{fp32} or \texttt{int32}). This type of mixed precision computation is often called \emph{quantization} for integer values~\cite{tflite_quant}. 
In this paper, we will always use \emph{mixed precision} for brevity.

\begin{figure}[t]
    \centering
    \includegraphics[width=0.9\linewidth]{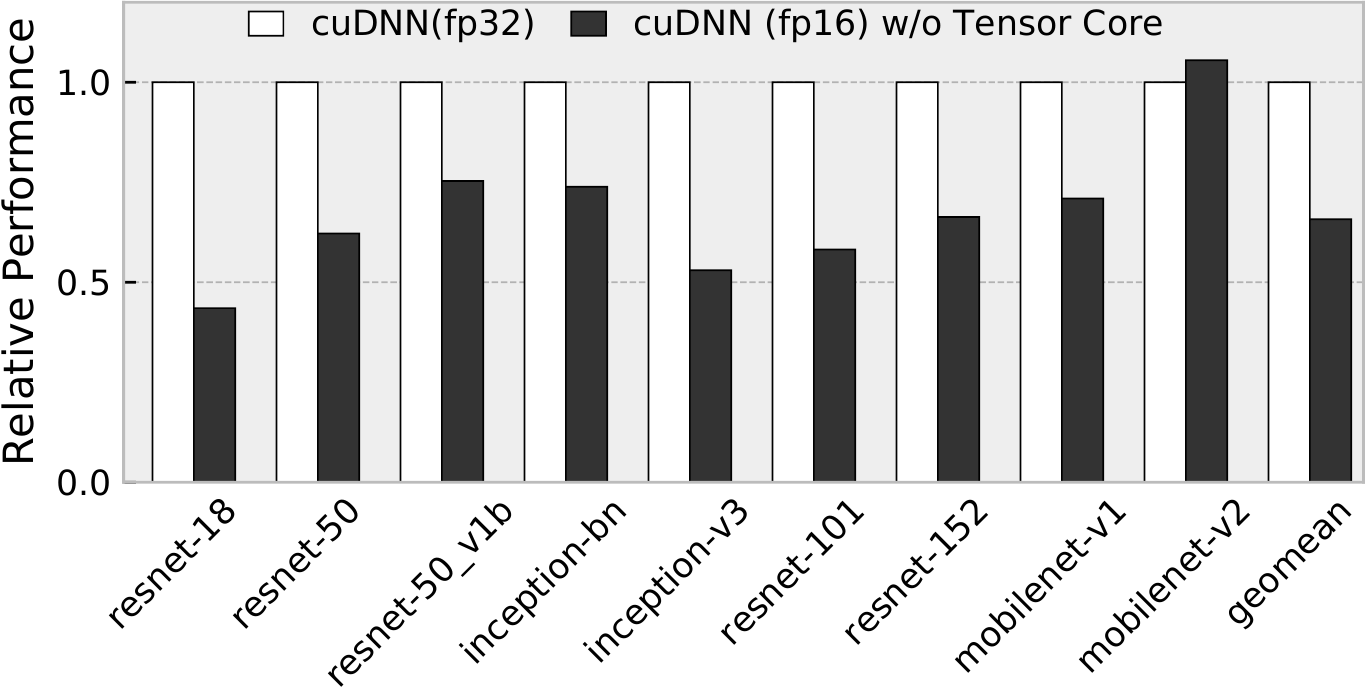}
    \vspace{-0.05in}
    \caption{Performance comparison on Nvidia V100-SXM2 between \texttt{fp32} and \texttt{fp16} without mixed precision instruction support.}
    \label{fig:bg-perf}
\end{figure}

While using mixed precision data types reduces memory footprint, it might not necessarily lead to performance improvement. To investigate this, we conducted an experiment to compare the performance of Nvidia cuDNN performance for \texttt{fp16} and \texttt{fp32} in the absence of Nvidia mixed precision tensorized instructions (Tensor Core). As shown in Figure~\ref{fig:bg-perf}, we observe that blindly using mixed precision leads to substantial slowdown because of the overhead of casting between two data types.

Therefore, mainstream hardware vendors (Intel, ARM and Nvidia) have introduced mixed precision tensorized instructions to achieve better performance. These instructions add mixed precision arithmetic support where operands are of lower precision while the accumulation happens in higher precision, potentially leading to 2$\times$ - 4$\times$ speedup. The most popular examples of these tensorized instructions are Intel VNNI, ARM DOT and Nvidia Tensor Core. We will discuss the semantics of these operations in Section~\ref{sec:tensorizer}.

\begin{figure}[t]
    \centering
    \includegraphics[width=0.9\linewidth]{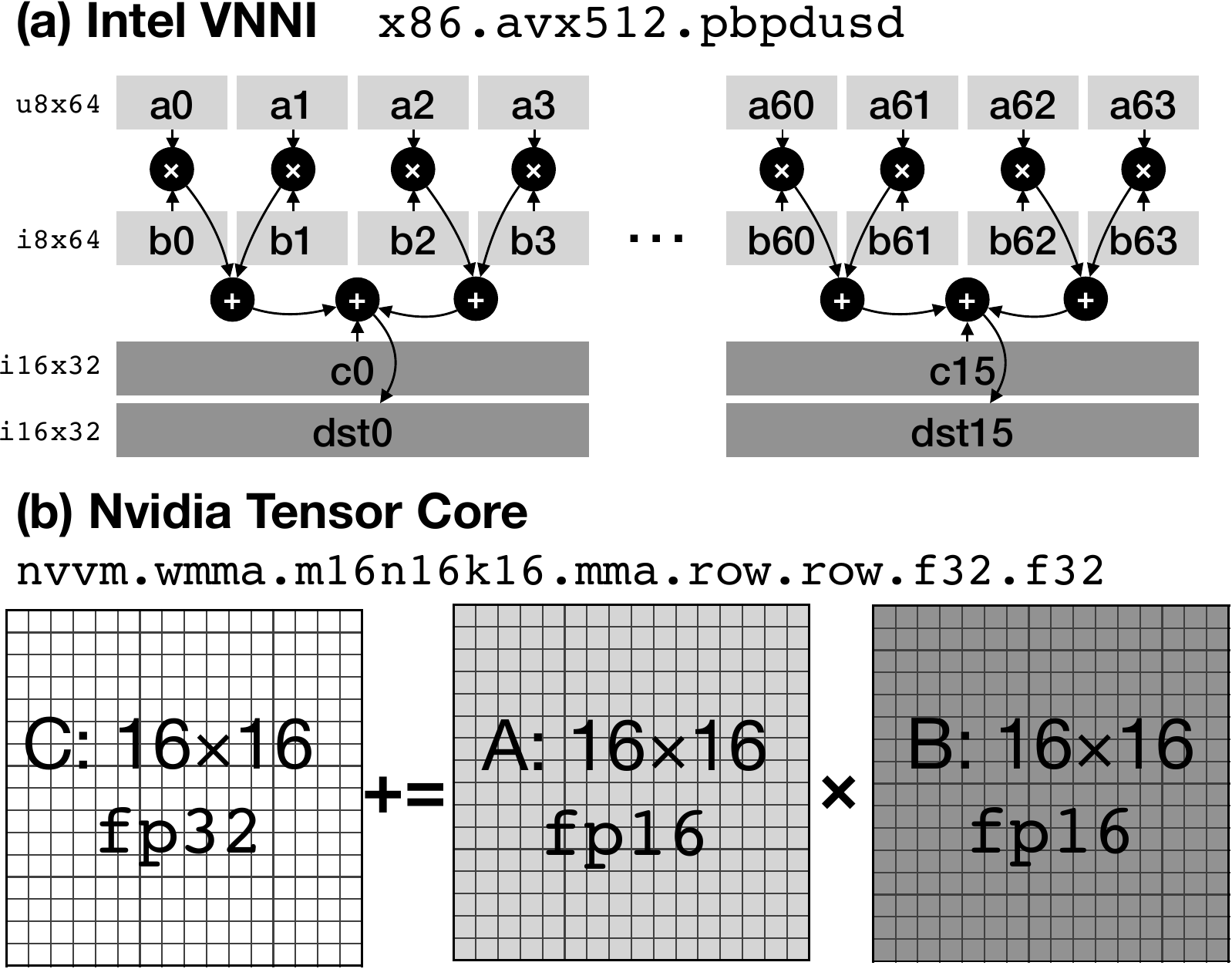}
    \vspace{-0.0in}
    \caption{The semantics of Intel VNNI and Nvidia Tensor Core. The text beside is the name of the corresponding LLVM intrinsic.}
    \label{fig:datapath}
\end{figure}

Hardware vendors have a long history of adding new instructions to accelerate important applications. However, the mixed precision tensorized instructions introduce a unique idiom - horizontal accumulation. These tensorized instructions typically conduct a sequence of elementwise multiplications governed by a memory access pattern, followed by a horizontal accumulation. The accumulation is termed horizontal because all values to be accumulated are present in the same vector register. For example, as it is shown in Figure~\ref{fig:datapath}(a), Intel VNNI executes a dot product of two vectors, each having 4 \texttt{int8} elements, while performing the accumulation in \texttt{int32}. We observe a similar pattern, though with different numbers of entries and data types, for Nvidia Tensor Core (in Figure~\ref{fig:datapath}(b)) and ARM DOT instructions (this is omitted, because it is similar to VNNI).

\subsection{Limitations of Existing Solutions}
Though tensorized instructions seem promising, their adoption pace is limited because of the absence of an automatic technique that can detect and use these instructions seamlessly. Currently, their usage in the deep learning domain is limited to hardware vendor libraries like Intel oneDNN and Nvidia cuDNN, which may provide high performance for the pre-defined operations but are inflexible as discussed in Section~\ref{sec:intro}.

Similarly, conventional loop vectorizers find it hard to exploit the profitability of these tensorized instructions, as they are not designed to work with the horizontal reduction idiom. Conventional loop vectorizers in general-purpose compilers like GCC and LLVM mainly focus on either analyzing the innermost loop body or combining instructions in the unrolled loop bodies. When it comes to the horizontal reduction idiom, these compilers often reorder the computation and generate epilogue reduction, preventing us from using the tensorized instructions.

There have been some recent works in compiling programs to leverage
tensorized instructions. PolyDL~\cite{polydl} generates CPU programs for convolution kernels in neural networks that call a GEMM micro-kernel using Intel VNNI instructions. Bhaskaracharya et al.~\cite{nv-auto-tc} generate CUDA programs for matrix computation leveraging Nvidia Tensor Core. However, these works are limited to one platform and its specific instruction, which lacks generalizability. A generic solution to handle tensorized instructions from multiple platforms together is still missing.


\subsection{Multi-Level Intermediate Representation}

Compilers often have multiple levels of intermediate representation (IR) to express the program; each level is designed to enable different analyses and transformations. In this section, we describe the background of a tensor domain specific language (DSL) and the multi-level IR.

\subsubsection{Graph-Level IR} Deep learning compilers like TVM~\cite{tvm}, Glow~\cite{glow}, and XLA~\cite{xla} adopt a graph-level IR to represent a deep learning model as a directed acyclic graph (DAG) of operations. This graph-level IR is useful for inter-tensor-operation optimization, like tensor shape padding, operation fusion, and choosing the proper data layout~\cite{graph-tuner}. Our tensorized analysis relies on tensor padding so that loops can be tiled by the number of lanes of the instruction perfectly. However, this IR has little knowledge about the implementation of each tensor operation. When compiling a graph-level IR, each node of the DAG will be dispatched to its  implementation in tensor DSL as explained next.

\subsubsection{Tensor DSL} 
Tensor domain-specific languages, like Halide~\cite{halide}, TVM~\cite{tvm}, and Tensor Comprehension~\cite{tensor-comprehension}, have been developed to productively and portably
express tensor programs while enabling efficient
performance tuning.  As shown in Figure~\ref{fig:sema} and Figure~\ref{fig:tensor-example}, programs written in tensor DSLs follow this paradigm: Users first declare the tensors and the loop variables, and then the computation is described by expressions involving the declared tensors and loop variables. These DSLs also provide interfaces to split, reorder, and annotate loops without affecting the computation semantics for performance tuning.

All the information gathered from the tensor DSL frontend will be stored in a \emph{tensor Op} data structure, including the declared tensors, loop variables, expressions, and loop manipulation.

\subsubsection{Tensor IR} Each \emph{tensor Op} is then lowered to \emph{Tensor IR}, which is an imperative program IR with additional constraints: All the loops are canonical (starting from 0, and increased by 1 each time), and all the array operations are restricted (i.e., an element cannot be accessed by two different pointers). These two properties enable
making strong assumptions for analysis and transformation. Our work conducts analysis on the \emph{tensor Op} data structure level and then performs transformation on the tensor IR. Although the tensor IR provides essentially identical information for analysis, as discussed above, it is easier to reorganize the loops via the \emph{tensor Op} data structure.

\subsubsection{Low-Level IR} The tensor IR is lowered to a general-purposed low-level IR like LLVM, after all the specialized analysis and transformations on the tensor IR are done, to get ready for assembly code generation.

\begin{figure}[t]
    \centering
    \includegraphics[width=0.8\linewidth]{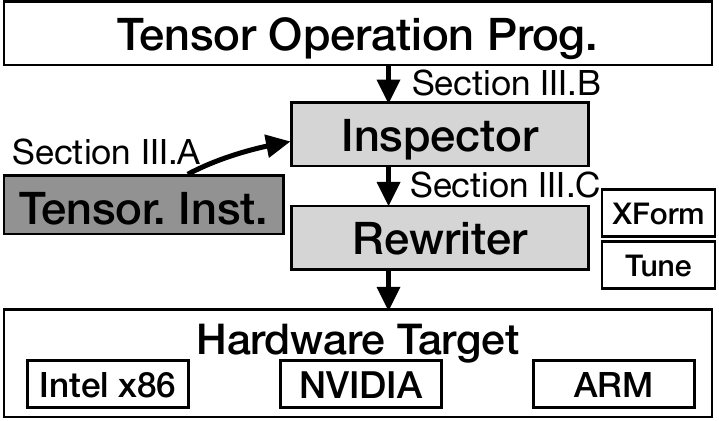}
    \vspace{-0.0in}
    \caption{The overview of our framework, \system.}
    \label{fig:pipeline}
    \vspace{0in}
\end{figure}

\section{\underline{Uni}fied \underline{T}ensorization} \label{sec:tensorizer}

Our goal is to automatically \emph{tensorize}\footnote{We coin the word to mean rewrite and optimize a given code by the tensorized instruction.} mixed-precision deep learning tensor operations across a variety of hardware platforms. We resolve the challenges discussed in Section~\ref{sec:intro} by presenting \system with the following techniques:
\begin{enumerate}
    \item \emph{Tensorized Instruction in Tensor DSL:} To abstract the diverse tensorized instructions on different hardware platforms, we leverage the existing tensor DSL to represent their semantics.
    \item \emph{Applicability Inspection:} To determine if and how a tensorized instruction can be applied to a tensor operation, we developed an analysis pass in the \emph{Inspector} component of \system, which analyzes the \emph{tensor Op} data structure of both the instruction and the operation. The result of analysis will guide the loop reorganization and instruction injection.
    \item \emph{Code \rewriter:} Once the tensorized instruction is determined applicable, the \rewriter reorganizes the loop nests in accordance with the Inspector so that the innermost loop nests resemble the tensorized instruction and are ready to be replaced. Finally, it sets up the tuning space for the remaining loop nests to exploit high performance.
\end{enumerate}


\ignore{
\begin{enumerate}
    \item \emph{Tensor DSL and Tensor-level IR:} We leverage existing tensor DSL to represent the semantics of tensorized instructions, and ubiquitous tensor-level IR to solve the mapping problem across a variety of hardware platforms.
    \item \emph{Inspector:} We develop a new tensor-level analysis IR pass - Inspector - to detect the applicability of a tensorized instruction. Inspector inspects the tensor-level IR of both tensor operations and tensorized instruction, and checks if we can map the instruction onto the operation, and also informs \rewriter (explained next) how loop nests should be organized to make the mapping easy.
    \item \emph{\rewriter:} \rewriter reorganizes the loop nests of the original tensor operation as per the inspection such that the innermost loop nests now resemble the tensorized instruction, and then replaces the innermost kernel with the tensorized instruction. Finally, \rewriter sets up the tuning space for the remaining loop nests, quickly tunes it on the hardware platform, generating a high performance implementation.
\end{enumerate}
}

These components of \system together enable a unified compilation flow to simplify the mapping of tensorized instructions across a variety of hardware platforms. In the rest of this section, the details of each of the above steps will be discussed.

\begin{figure}[t]
    \centering
    \includegraphics[width=\linewidth]{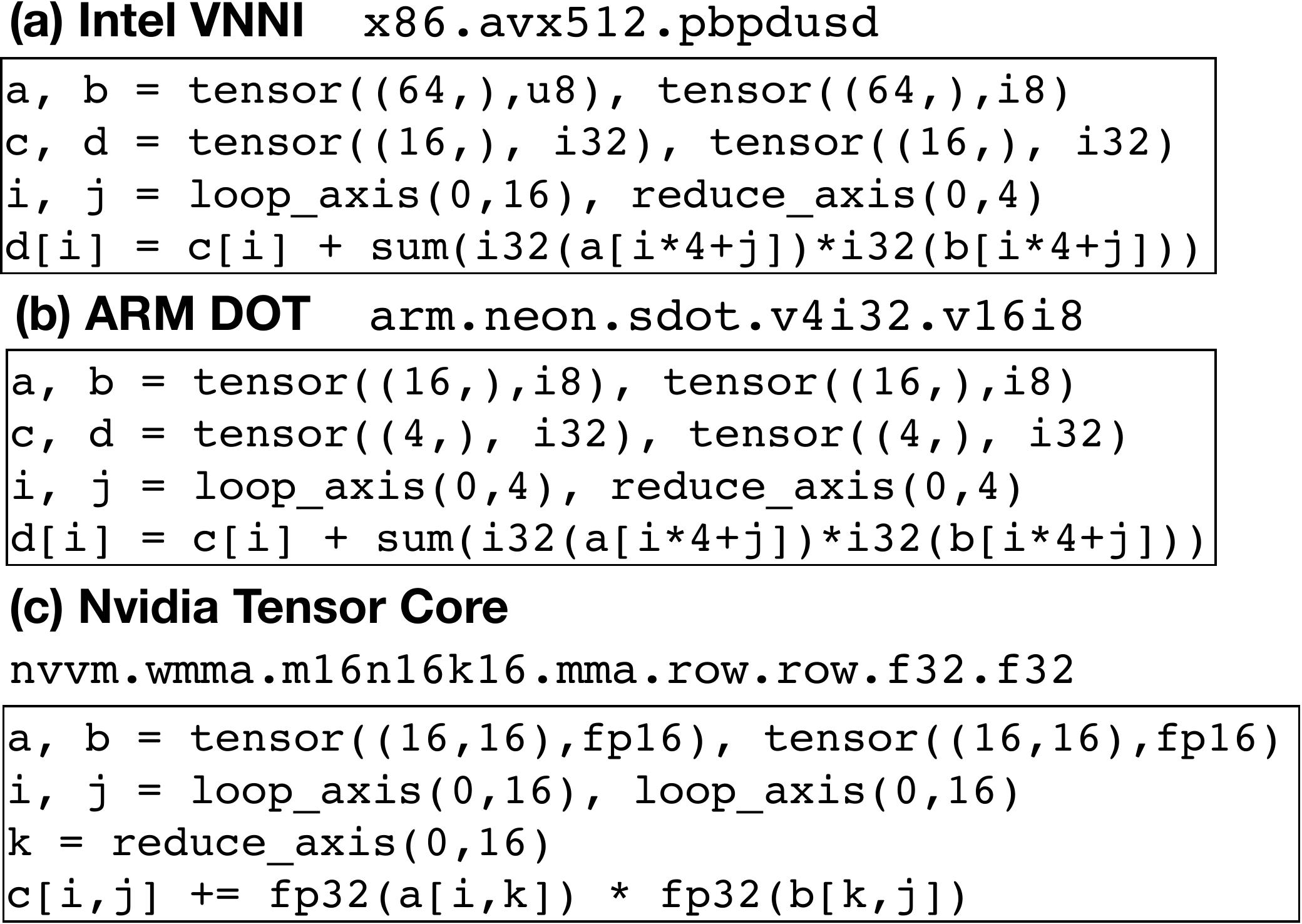}
    \caption{Tensorized instructions as abstracted in the tensor DSL.}
    \label{fig:sema}
\end{figure}

\subsection{Semantics Abstraction - Tensor DSL}
In order to unify the compilation of tensorized instructions from different platforms and keep the system open to integrate new instructions, the first question to answer is how to have a unified description of the semantics of tensorized instructions. As explained in Section~\ref{sec:bg}, we employ ubiquitous tensor DSL and tensor IR to solve the abstraction problem. All mixed precision tensorized instructions perform some elementwise operations for vectors, followed by a horizontal reduction. Each tensorized instruction, therefore, can be regarded as a small tensor operation program written in the tensor DSL.

\begin{figure}
    \centering
    \includegraphics[width=0.88\linewidth]{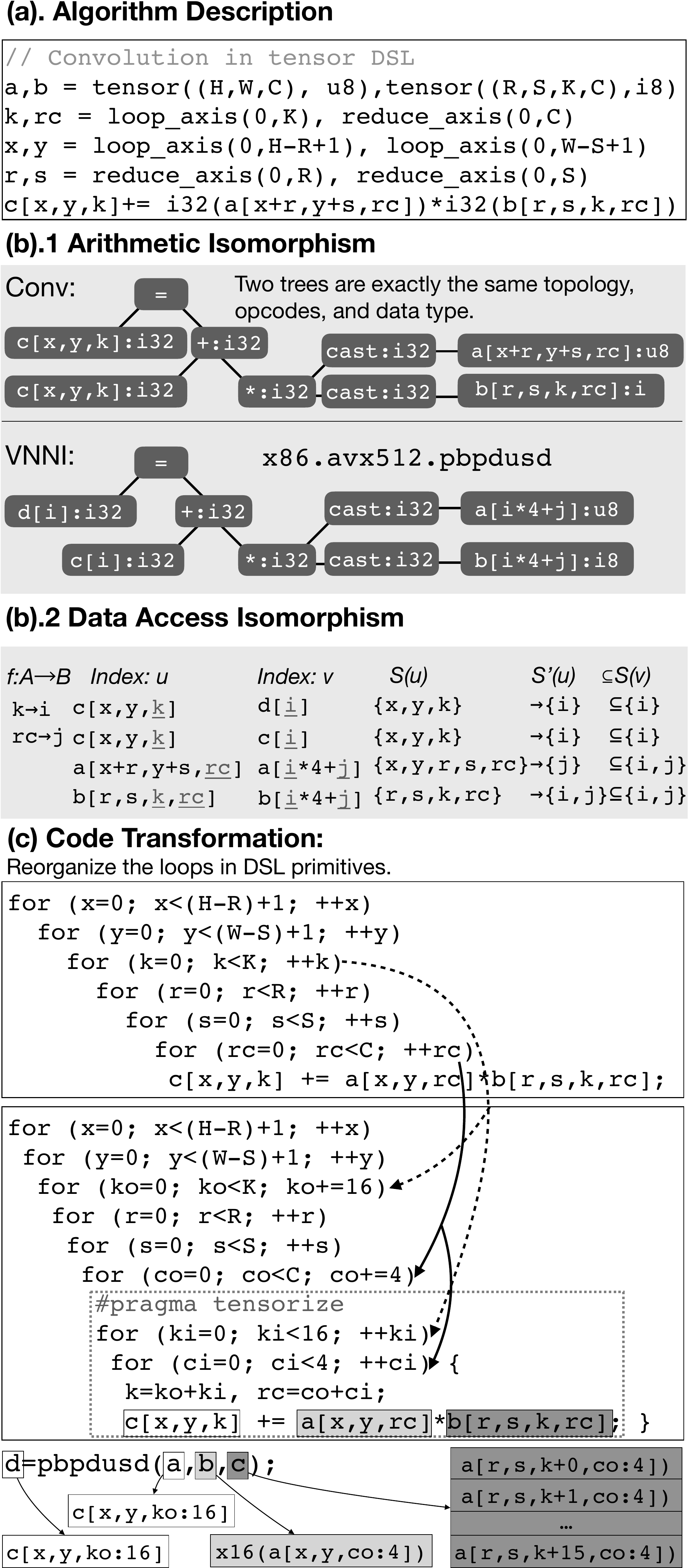}
    \caption{An example of applying Intel VNNI to Conv using \system.}
    \label{fig:tensor-example}
\end{figure}

Figure~\ref{fig:sema}(a) shows how an Intel VNNI instruction is described in the tensor DSL. Three source operands of Intel VNNI are 512-bit registers. Two of them are 64 lanes of unsigned 8-bit integers (\texttt{uint8}) and signed 8-bit integers (\texttt{int8}), and the other one is 16 lanes of signed 32-bit integers (\texttt{int32}), which correspond to the tensors \texttt{a}, \texttt{b}, \texttt{c} we defined. The arithmetic behavior is defined by the loop variables and the expression of \texttt{d[i]}. Here we annotate that loop \texttt{i} is data parallel, since these 16 elements are independent from each other; loop \texttt{j} is reduction since for every independent element it sums up 4 elements along with this loop. A similar loop pattern appears in the other tensor operations shown in Figure~{\ref{fig:tensor-example}}. The description of ARM DOT, shown in Figure~\ref{fig:sema}(b), is similar to Intel VNNI, with a different number of lanes and data types.

Nvidia Tensor Core, on the other hand, performs a $16^3$ square matrix multiplication as shown in Figure~\ref{fig:sema}(c). Comparing with (a) and (b), a key difference is that it requires the accumulator register to be the same as the addition register (note the \texttt{+=}). This is due to the data type opaqueness of the Tensor Core instruction, which prevents us from giving arbitrary initial values for the accumulators.

We describe the semantics of each tensorized instruction in tensor DSL. The deep learning compiler pipeline parses the operation into \emph{tensor Op}, which preserves tensor information like the expression tree, the loop trip count, and the array buffers. This information is essential for the analysis and transformation passes in Inspector and \rewriter.

\subsection{Applicability Detection - Inspector} \label{subsec:app}
To determine if a tensorized instruction can be applied to a tensor operation, the Inspector pass uses a two-step approach. It first determines if (part of) the tensor operation program and the instruction can be arithmetically equivalent by checking a form of isomorphism between their associated expression trees. After that, it inspects the data access pattern to confirm the assembly operands can be prepared so as to guide the \rewriter transformation.

\begin{algorithm}[t]
{\footnotesize \caption{Determine the isomorphism between expression trees. \emph{a} is for the instruction, and \emph{b} is for the operation.\vspace{0.1in}} \label{algo:iso}}
\begin{algorithmic}
\vspace{0.1in}
\Function{Inspect}{a,b}
  \If{a.type=b.type}
    \If{isleaf(a)$\land$isleaf(b)}
      \If {a is not bound}
        \State bind[a]:=b
      \ElsIf{bind[a]$\ne$b}
        \State \textbf{return} False
      \EndIf
      \State \textbf{return} True
    \ElsIf{isarith(a), isarith(b)}
       \State cond:=a.opcode=b.opcode
       \State cond:=cond$\land$Inspect(a.lhs, b.lhs)
       \State cond:=cond$\land$Inspect(a.rhs, b.rhs)
       \State \textbf{return} cond
    \EndIf
  \EndIf
  \State \textbf{return} False
\EndFunction
\end{algorithmic}
\end{algorithm}

\subsubsection{Compute Isomorphism} Algorithm~\ref{algo:iso} shows the algorithm we adopt to determine the isomorphism of two expression trees. It recursively traverses both trees and matches the data type and opcode of each pair of nodes. Figure~\ref{fig:tensor-example}(b).1 shows that the two trees of convolution and \texttt{pbpdusd} (an Intel VNNI instruction) are in exactly the same topology and data type, so these two programs are arithmetically isomorphic.

This analysis also finds a mapping from the operands in the tensor program to the operands in the tensorized instruction. As we explained, tensor operands in the tensorized instruction are the abstraction for registers. Therefore, a register cannot correspond to multiple data sources. This property still requires further checks, which will be explained in the next section.

\subsubsection{Array Access Isomorphism} Once compute isomorphism is determined, the next concern is how the data are fed to this instruction. The enforcement explained in the last subsection already determines each register operand only corresponds to one array in the tensor operation. On top of this, we need to determine each element in the operand tensor corresponds to only one memory address in the tensor program when mapping to the tensorized instruction. 
To map a tensor program to a tensorized instruction, we need to know which loop levels are tensorized. We enumerate the loop levels to be tensorized, and these loop levels will be mapped to loops in the tensorized instruction. Note that only loops with the same annotation (data parallel or reduction) can be mapped to each other. Then we check if this enumerated mapping is feasible, by scanning each pair of operand correspondence determined in the last paragraph. If the operand in the tensor program is a constant, we just skip it\footnote{If it is a constant, the correspondence was already checked in the last section. This register corresponds to this constant.}. If the operand is a memory operation, we inspect the index expressions of both memory operations in the operation and instruction. We define:
\begin{itemize}
    \item $A$ is the set of loop variables to be mapped to the tensorized instruction.
    \item $B$ is the set of loop variables of the tensorized instruction.
    \item $f: A\mapsto B$ is the mapping we enumerate.
    \item $S(u):=\{x|x\text{ is loop variable in the index expression }u\}$
    \item $S'(u):=\{f(x)|x\in S(u)\cap A \}$
\end{itemize}

A mapping is considered feasible, if every pair of memory operation's index expressions $(u,v)$, where $u$ is from the operation and $v$ is from the instruction, holds $S'(u)\subseteq S(v)$. Figure~\ref{fig:tensor-example}(b).2 shows an example of inspection. If $S'(u)$ is a subset of $S(v)$, this means the data loaded by the tensor operation should be broadcast along with the loop variables that do not exist in $S(v)$ to fill all the register lanes. If not, this means each register lane corresponds to multiple memory addresses under this mapping, which is not realistic for code generation, so we should try another enumeration.

If there are multiple feasible mappings, we leave this as a dimension of code tuning space. Once this mapping is determined, it will guide the further loop transformation and code generation.


\subsection{Code Transformation - \rewriter} \label{subsec:rewrite}

There are three phases in the code transformation: loop reorganization, tensorized instruction replacement, and tuning.

\subsubsection{Loop Reorganization} As discussed in Subsection~\ref{subsec:app}, the inspector selects the loop levels to be executed by the given instruction. To get poised for code generation, as shown in Figure~\ref{fig:tensor-example}(c), we need to tile these loops and reorder them to the innermost loop levels so that those innermost loops perform exactly the same semantics as the instruction. As we explained, tensor DSL provides the capability to reorganize the loops nests easily.

\ignore{
the tensor DSL decouples the computation and code organization\yida{decouples computation and scheduling?}, so these can easily be achieved by calling loop organization interfaces. }

\subsubsection{Tensorized Instruction Replacement} After identifying the code region to be replaced by a tensorized instruction, the code generator should prepare each operand of this instruction. It is difficult to fully automate the operand preparation for different platforms because of their diverse execution models and assembly formats. Therefore, we formalize a unified programming interface to compiler developers to manually specify the rule of operand generation. In this interface, each loop variable to be replaced, and their coefficients in the index expression are exposed. For example, as shown in Figure~\ref{fig:tensor-example}(c), by analyzing the strides and trip count of \texttt{ki}, and \texttt{ci}, the array access \texttt{c[x,y,c]} will be transformed to a 16-lane vector; \texttt{a[x,y,rc]} will be vectorized along with \texttt{c} by 4, and broadcast along with \texttt{ki} by 16; \texttt{b[r,s,k,c]} will be vectorized along with \texttt{ci} by 4, and unrolled and concatenated along with \texttt{ki}.

\subsubsection{Tuner} All the other loop levels that are not involved in instruction rewriting can be reorganized to tune the performance. Here, we develop strategies to optimize the performance of tensor programs on both CPU and GPU. The generic philosophy is to exploit both fine- and coarse-grained parallelism. We also developed specialized strategies because of the different execution models and memory hierarchy.

\noindent \textbf{CPU Tuning:} On CPU, data-parallel loops are distributed to multiple threads to achieve coarse-grained parallelism. On the other hand, the loop-carried dependence in reduction loops introduces RAW hazards in the execution pipeline. To avoid this penalty, and achieve instruction-level parallelism, we reorder and unroll a small degree of data parallel loops below the innermost reduction loop.

The tuning space of CPU involves two dimensions, the degree of unrolling and parallelization. We enumerate these two parameters and profile the execution time to search for the best one. If the unrolling degree is too small, there will not be enough independent instructions to fill in the idle penalty cycles caused by RAW hazards. If it is too large, it will cause I-cache misses. Similarly, the number of threads can neither be too few or too many. If it is too few, the computing cores would have insufficient utilization and memory latency would not be hidden. Too many threads introduce context switching overhead. We rely on the tuning process to look for the best combination.

\begin{figure}[t]
    \centering
    \includegraphics[width=0.85\linewidth]{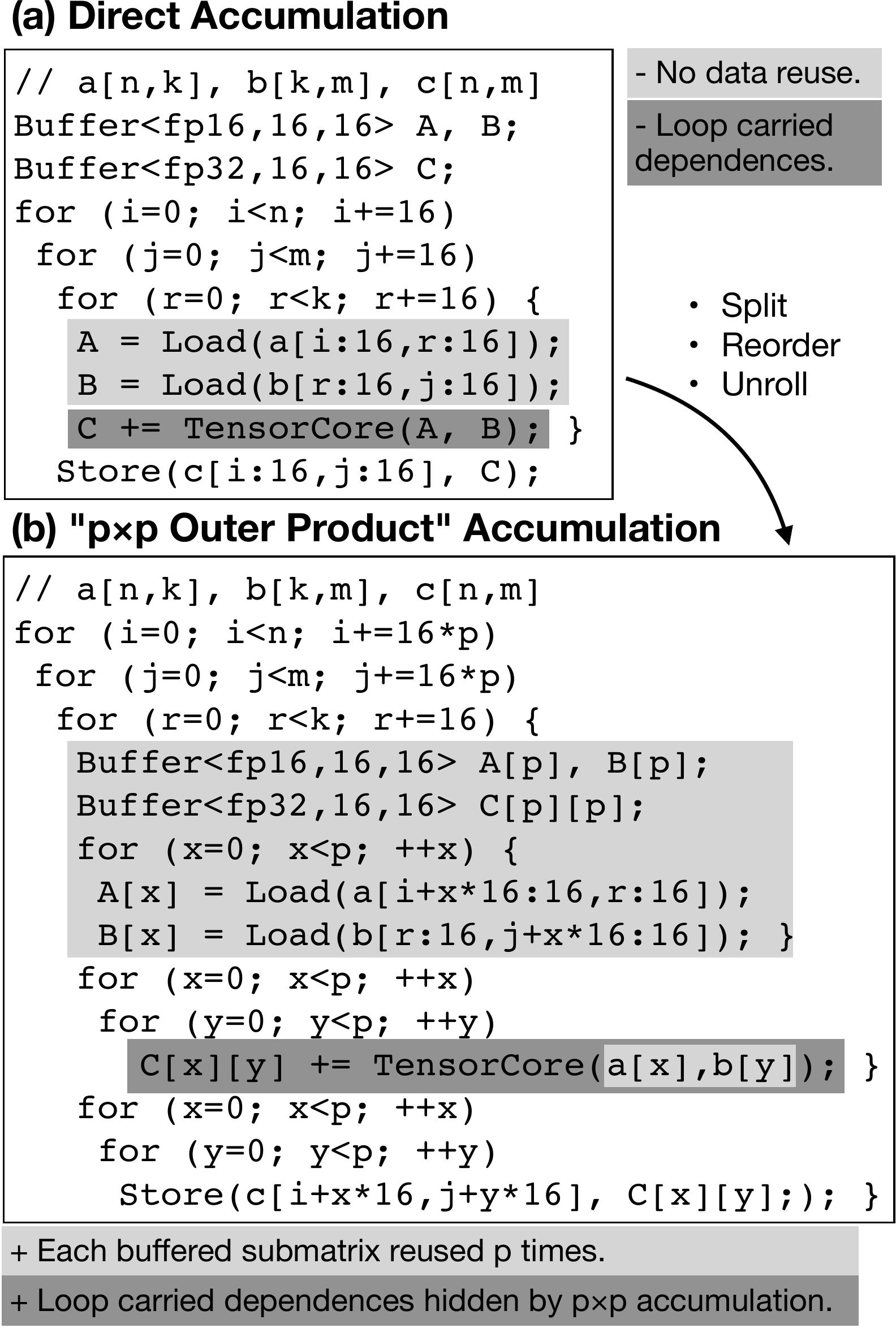}
    \caption{Accumulating a p$\times$p ``square window'' avoids loop-carried data dependences, and reuses buffered submatrices.}
    \label{fig:gpu-out}
\end{figure}

\noindent \textbf{GPU Tuning:} On GPU, coarse-grained parallelism is achieved by distributing the data parallel loops across the streaming multiprocessors. Similar to CPU,  fine-grained parallelism is also achieved by reordering and unrolling a small degree of data parallel loops to avoid the pipeline penalty caused by loop-carried dependences. Moreover, on GPU, data reuse is explicitly managed by the software. Therefore, as it is shown in Figure~\ref{fig:gpu-out}, we adopt an outer-product style matrix multiply accumulation to reuse the buffered submatrices.

Besides the generic optimization, we also developed optimization mechanisms specialized for DNN kernels. Among popular DNN models, there are many layers with relatively small width and height and deep channels. We apply \emph{dimension fusion} to layers with small width and height -- these two dimensions are fused into one to save the redundant padding. In addition, we apply \emph{split reduction} to layers with deep channels. For a reduction loop with large trip count, we can split it and parallelize each split segment on \texttt{threadIdx}. After all the segments are done, we synchronize the threads and reduce the splitted segments in the shared memory.

\vspace{0.05in}
\section{Implementation} 
\label{sec:impl}
\vspace{0.05in}

In this section, we will discuss technical details in our implementation. \system is implemented by extending Apache TVM~\cite{tvm}, a full-stack deep learning compiler, with tensorized instruction support. We leverage TVM's tensor DSL, tensor Op, tensor IR infrastructure, and the tuning infrastructure mechanisms~\cite{autotvm, graph-tuner} to generate high performance kernels. In addition, implementing \system on top of TVM enables end-to-end model inference with other optimizations such as operator fusion, in addition to tensorization.

\ignore{
\subsection{Analysis and Transformation}
}

\subsection{Inspector}
The inspector pass is implemented by analyzing TVM's \texttt{ComputeOp} data structure. This matches the expression tree of both the instruction and program and enumerates mappings between the loop variables. We enumerate the loops from the tensor's innermost dimension to outermost dimension, and greedily return the first eligible one because of the better potential data locality for inner dimensions. The enumerated mapping provides us with the correspondence of loop variables between the instructions and the tensor operations.

\subsection{\rewriter}
These following steps will be performed by the rewriter:
\begin{enumerate}
    \item According to the loop correspondence analyzed by the inspector, we reorganize the loops to be tensorized by tiling these loops by the trip counts of the corresponding loops in the instruction, and reorder them to be the innermost loops. These loops will be annotated by a \texttt{tensorize} pragma to hint the instruction injection.
    \item Based on the strategies discussed in Section~\ref{subsec:rewrite}, we reorganize the loops above not involved in instruction rewriting to tune the performance.
    \item We lower the manipulated loop nest to the tensor IR, and replace the loop body annotated with the \texttt{tensorize} pragma with the target instructions, as shown in Figure~\ref{fig:tensor-example}(c).
\end{enumerate}

Steps 1 and 2 are achieved by invoking TVM scheduling primitives on the tensor DSL level, and step 3 is a tensor IR transformation pass. 

\begin{figure}[t]
    \centering
    \includegraphics[width=0.83\linewidth]{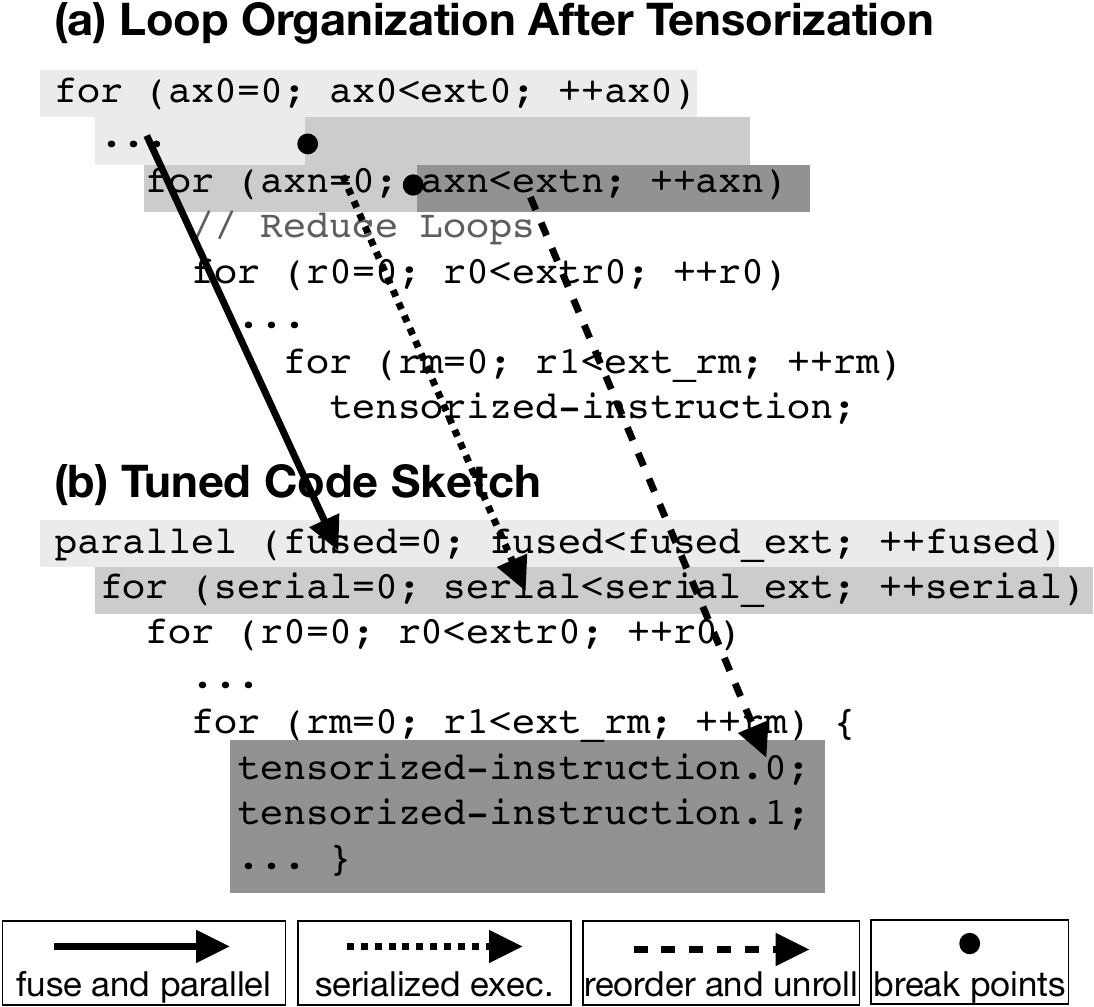}
    \caption{The code sketch of CPU tuning.}
    \label{fig:cpu-tune}
\end{figure}

Next, we discuss the implementation of the tuning strategies discussed in the last section.

\noindent \textbf{CPU Tuning:} The code sketch of tuned CPU code is shown in Figure~\ref{fig:cpu-tune}. To implement the tuning we discussed in Section~\ref{subsec:rewrite}, we enumerate two breaking points on the data parallel loop nest, which define how the loop levels are parallelized and unrolled.  
A breaking point is defined by a \emph{loop level} and \emph{tiling factor}, giving more flexibility to the division.
Loops before the first breaking point, will be fused and parallelized. Loops between these two points will be executed in serialized order. 
Loops after the second breaking point will be reordered to the innermost and unrolled.  


\noindent \textbf{GPU Tuning:} As it is discussed in the last paragraph of Section~\ref{subsec:rewrite}, both coarse-grained and fine-grained parallelism optimizations are applied on data-parallel loops, so there is a tradeoff between them: data reuse is increased by increasing the unrolling degree (each buffered submatrix is reused \texttt{p} times), but the coarse-grained parallelism is decreased. Also, a large unrolling degree may overwhelm the register resources. Therefore, the key to generic optimization is to choose a proper unrolling degree.

On the other hand, greedily applying each specialized optimization does not always improve the performance. Though dimension fusion may save the memory traffic, it also introduces software overhead on data rearrangement. Similarly, though splitting the reduction loop introduces more parallelism, it also introduces thread synchronization overhead and register pressure.  We enumerate each parameter, including the degree of reduction parallelization and whether to fuse the width and height dimensions, and then apply these transformations to the program and profile the performance to determine which transformation leads to the best performance.

\begin{figure*}[t]
\begin{minipage}{0.49\linewidth}
    \centering
    \includegraphics[width=0.9\linewidth]{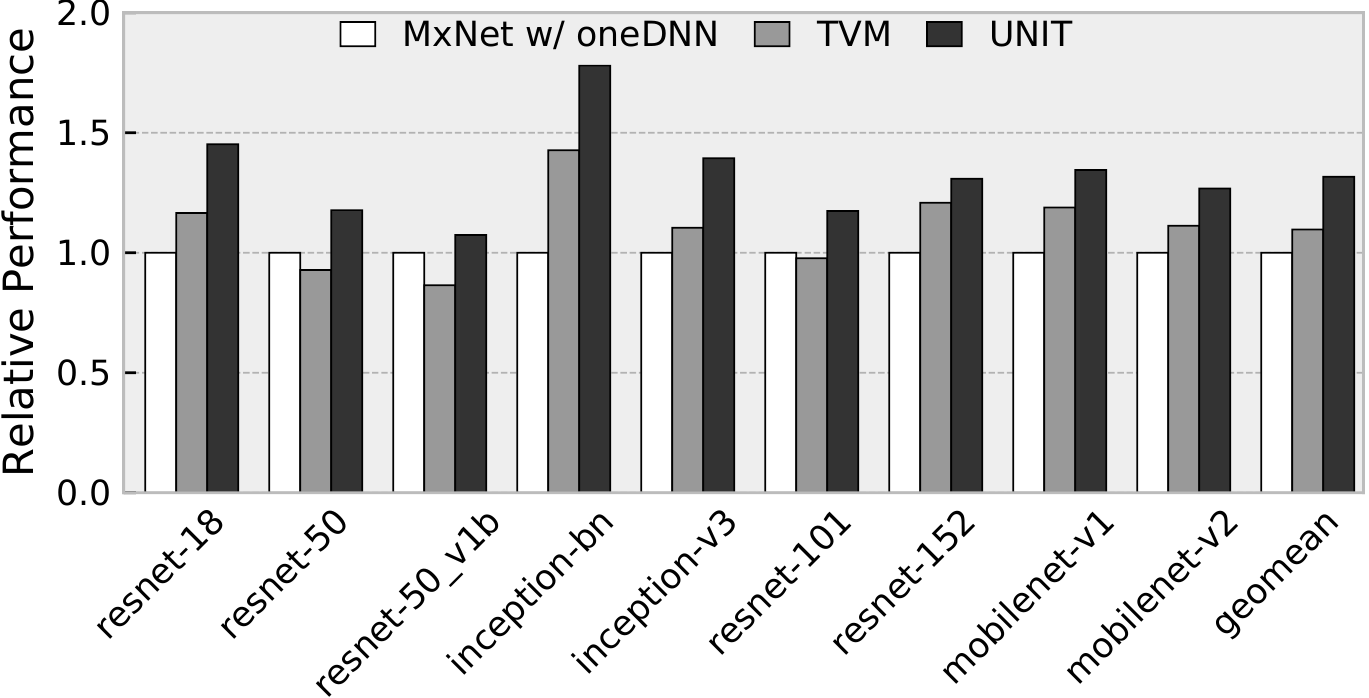}
    \vspace{-0.05in}
    \captionof{figure}{Quantized network inference (bs=1) accelerated by Intel VNNI.}
    \label{fig:e2e-cpu}
    \vspace{-0.25in}
\end{minipage}
\hfill
\begin{minipage}{0.49\linewidth}
    \centering
    \includegraphics[width=0.9\linewidth]{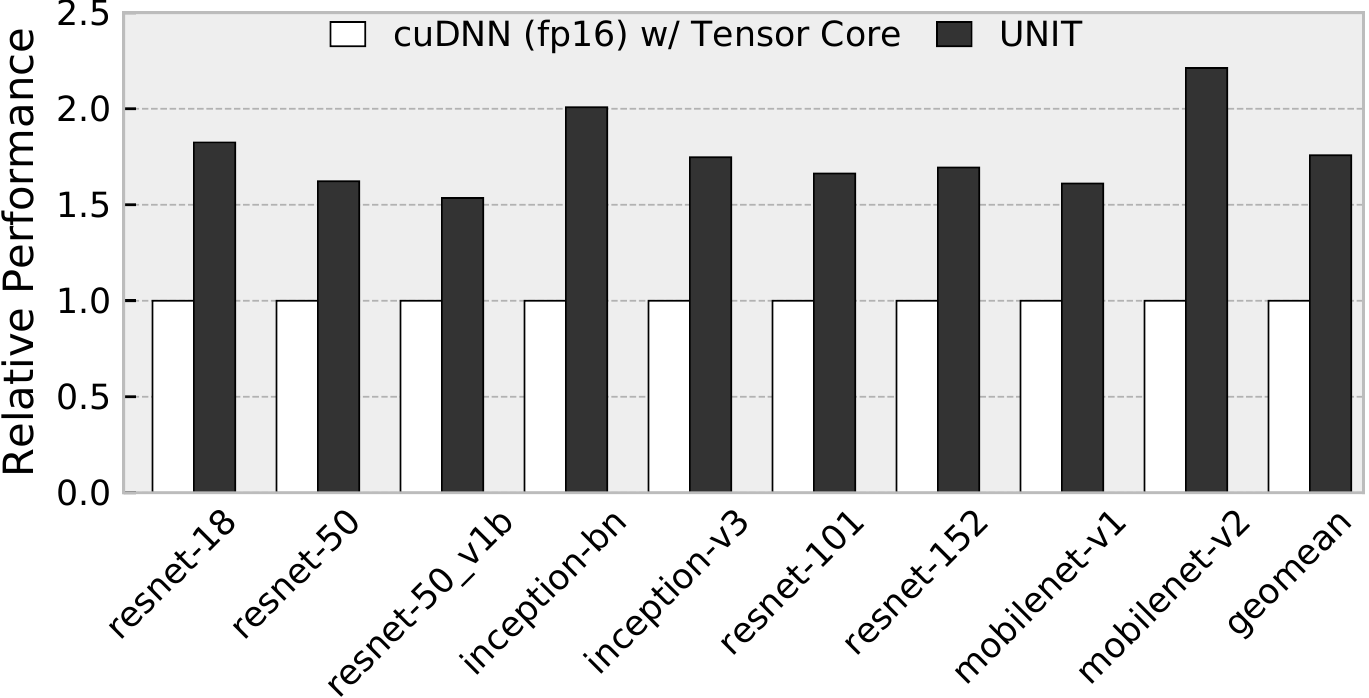}
    \vspace{-0.05in}
    \captionof{figure}{Mixed precision network inference (bs=1) accelerated by Tensor Core.}
    \label{fig:e2e-gpu}
    \vspace{-0.25in}
\end{minipage}
\end{figure*}

\vspace{0.1in}
\section{Methodology} \label{sec:method}
\vspace{0.1in}

\subsection{Target Hardware Platforms}

We assess \system on three hardware platforms:

\noindent \textbf{Intel x86 CPU:} We use Amazon EC2 C5.12xlarge instance as our x86 platform with 24-core Intel Xeon Platinum 8275CL CPU @3.00GHz (codename: Cascade Lake) and 96GB memory.

\noindent \textbf{ARM CPU:} We use Amazon EC2 M6g.8xlarge instance as our ARM platform with AWS Graviton2 CPU, which features 32-core ARM Cortex-A72 CPU @2.30GHz and 128GB memory.

\noindent \textbf{Nvidia GPU:} We use Amazon EC2 P3.2xlarge instance as our GPU platform with Nvidia Tesla V100 SXM2 GPU that has 16GB host memory.

\subsection{Software Frameworks}

\noindent \textbf{Code Generation:} All programs implemented in Apache TVM are emitted to LLVM IR for code generation. We choose LLVM-10 as our backend, and to be compatible, we use CUDA-10.0 as the NVPTX linker and runtime.

\noindent \textbf{Baseline:} We use vendor-provided libraries for baseline performance of operators whenever possible. Specifically, Intel oneDNN v1.6.1 and Nvidia cuDNN 7.6.5 are used as our CPU and GPU baselines, respectively. For end-to-end model inference, we looked for the best available solutions with those libraries, which was MXNet integrated with oneDNN for CPU and TVM integrated with cuDNN for GPU.
Another set of baselines is the manually written implementation. To this end, we use the existing TVM solutions for Intel and ARM CPUs, which involve heavy engineering effort to carefully write intrinsics to use Intel VNNI and ARM DOT instructions. We did not find a manually written Tensor Core implementation that covers our evaluated workloads.

\subsection{Workloads}
\noindent \textbf{DNN Models:} All DNN models are from the MXNet Model Zoo and converted to TVM's graph IR, Relay~\cite{relay}, for quantization~\cite{qnn}, layout transformation, and data padding. All these models adopt \texttt{NCHW[x]c} data layout \cite{graph-tuner} for the data and \texttt{KCRS[y]k[x]c} for the kernel. Here \texttt{N} denotes the batch size, \texttt{C} denotes the input channels, \texttt{H} and \texttt{W} are the width and height of the input image, and \texttt{[x]c} denotes that the original \texttt{C} is split by \texttt{x}. Similarly, \texttt{K} denotes the number of output channels, \texttt{R} and \texttt{S} are the height and width of the kernel, and \texttt{[y]k} denotes the original dimension \texttt{K} is split by \texttt{y}.  \texttt{[x]} equals to the number of lanes of the instruction output, and \texttt{[y]} equals to the width of reduction.

In the evaluation, we target the \texttt{N=1} cases, because it is hard to optimize but critical for inference use cases. Comparing with batched cases where \texttt{N>1}, we cannot reuse the kernel tensor across samples, or exploit the parallelism brought by the data-parallel batching dimension.

\section{Evaluation} \label{sec:eval}

Our evaluation of \system attempts to answer these questions:
\begin{enumerate}
    \item What is the performance of the end-to-end deep learning model inference powered by \emph{tensorized} instructions?
    \item How does each optimization technique that \system uses impact the performance?
    \item Can \system be extended to support new hardware platforms and tensor operations?
\end{enumerate}

\vspace{-0.1in}
\subsection{End-to-End Performance} \label{subsec:perf}

In this subsection, we show the \system end-to-end effectiveness on Intel x86 and Nvidia GPU processors for tensorizing mixed precision instructions. For Intel x86 experiments, we use MXNet integrated with Intel oneDNN (referred to as MXNet-oneDNN) as the baseline. Another comparison of ours is TVM with manually written schedules using Intel's VNNI instruction. The findings of this experiment are shown in Figure~\ref{fig:e2e-cpu}.

We observe that \system achieves significant speedup compared to MXNet-oneDNN. Note that Intel oneDNN has access to manually written schedules that have been aggressively optimized and tuned by domain experts. We also observe that TVM overall achieves better performance than MXNet-oneDNN, but has suboptimal performance on resnet50 and resnet50b, which were heavily tuned by oneDNN engineers. On the other hand, \system outperforms both baselines, by 1.3$\times$ over MXNet-oneDNN and by 1.18$\times$ over TVM.

Next, we test the efficacy of \system on utilizing Nvidia Tensor Core instructions for Nvidia GPUs. For the baseline, we integrate TVM with cuDNN, which has access to manually written aggressively tuned Tensor Core schedules. The findings of this experiment are shown in Figure~\ref{fig:e2e-gpu}. We observe that \system consistently achieves better performance than cuDNN with a mean speedup of 1.75$\times$ and up to 2.2$\times$.



\subsection{Optimization Implications} \label{subsec:ablation}
In this subsection, we focus on the convolution operators of the DNN models to perform an in-depth analysis of the impact of different optimization techniques used by \system's \rewriter. This is essentially an ablation study, showing how important different parts of \system are. There are 148 different convolution workloads (i.e., convolution with different feature map sizes, kernel sizes, strides, etc.) in the models, out of which we choose 16 representative convolution layers. These kernels cover diverse input shapes and strides. Other workloads behave similarly in the ablation study.
We summarize the characteristics, namely, convolution attributes, like shapes, strides, etc., of the selected workloads in Table~\ref{tab:conv-shape}.

\begin{figure}[t]
    \centering
    \includegraphics[width=0.95\linewidth]{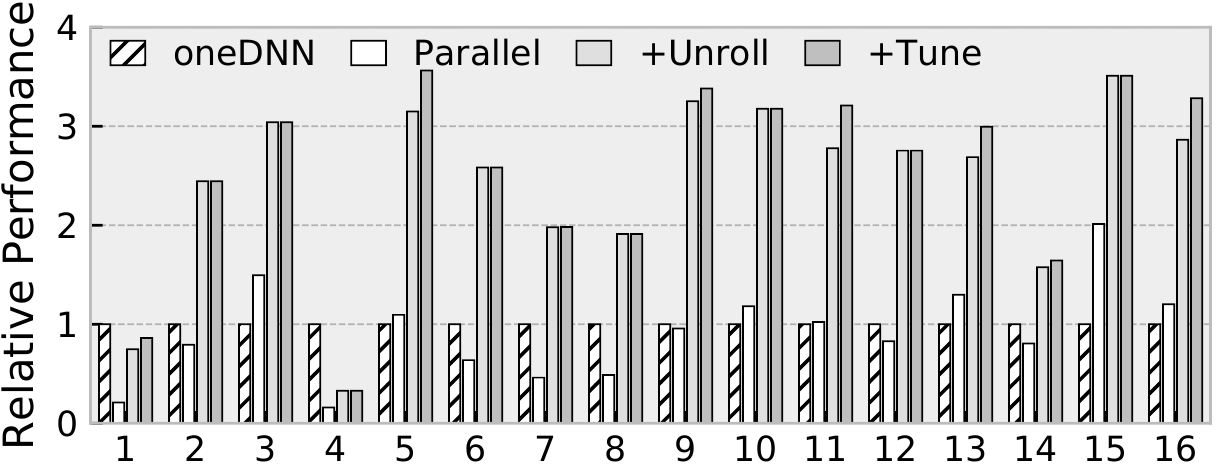}
    \vspace{-0.05in}
    \captionof{figure}{The performance impact of the code space exploration.}
    \label{fig:cpu-dse}
    \vspace{0.05in}
    \centering
    \includegraphics[width=0.95\linewidth]{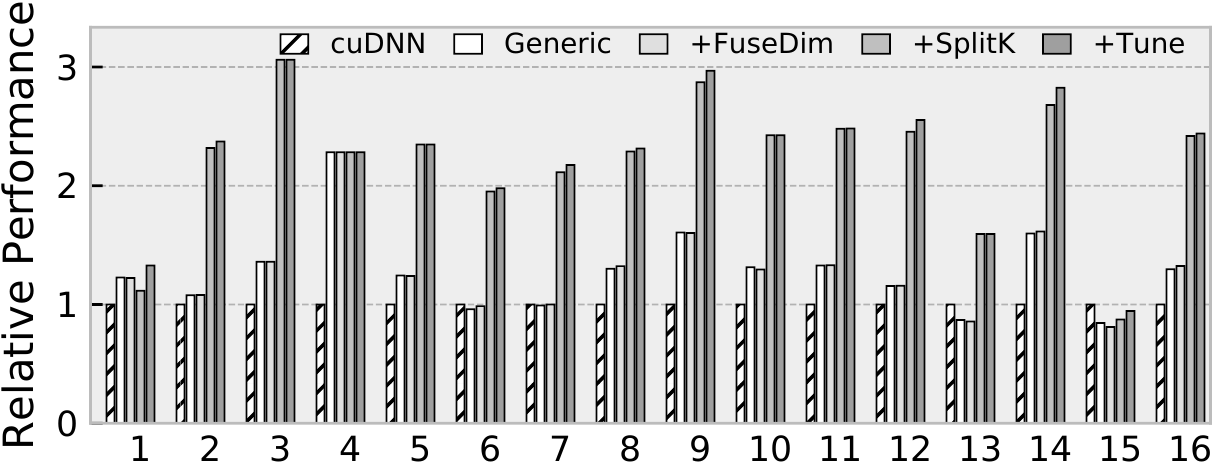}
    \vspace{-0.05in}
    \captionof{figure}{The performance impact of the code space exploration.}
    \vspace{0in}
    \label{fig:gpu-dse}
\end{figure}

\noindent \textbf{Intel x86 servers:} As we discussed in Section~\ref{subsec:rewrite}, we have two breaking points in CPU scheduling. The loop nests before the first breaking point are parallelized and the loop nests after the second breaking point are unrolled, while the ones in between the breaking point are executed serially. As loop nests can either be parallelized or unrolled (remaining one is serialized), we have a search space represented by the tuning pairs. \rewriter tunes this search space to generate a high-performance kernel. In this experiment, we incrementally measure the performance improvements brought by parallelizing, unrolling and tuning. The findings of this experiment are shown in Figure~\ref{fig:cpu-dse}, normalizing the speedup to Intel oneDNN execution latency.

First we fuse outer loop nests such that the loop bound of the fused loop nest is $<$ 3000, and measure the latency of the resulting kernel (shown by \emph{Parallel}). Then, we take the remaining loop nests, and tile and unroll them such the unrolling factor is $<$ 8, and measure this performance (shown by \emph{+Unroll}). Finally, instead of setting the limits as 3000 and 8, we tune the search space and measure performance (shown by \emph{+Tune}), getting the final latency \system achieves. We observe that Parallel and Unroll together is responsible for most of the speedup. The additional speedup introduced by Tuning is quite small. It turns out that more than half of the kernels get the optimal performance on the first tuning pair (i.e. 3000 and 8), and more than 95\% of the kernels get the optimal performance within the first 8 tuning pairs.

CPU does poorly on workloads \#1 and \#4, because their output shapes (OH/OW) can neither be perfectly tiled nor fully unrolled. Inherited from TVM, loop residues are handled the by guarding it with a likely clause, which results in an if-branch that harms the performance.

\noindent \textbf{Nvidia GPU servers:}  As discussed in Section~\ref{subsec:rewrite}, we employ three optimizations on GPU: generic coarse- and fine-grained parallelism, fusing width and height to save memory bandwidth, and parallelizing the reduction dimension. In this subsection, we study the impact of these optimizations on the performance. We show the findings in Figure~\ref{fig:gpu-dse}, normalizing the speedup to Nvidia cuDNN.

According to our evaluation, any unrolling degree (\texttt{p} in Figure~\ref{fig:gpu-out}) larger than 2 may overwhelm the registers, so we use \texttt{p=2} to apply the generic optimization. The generic optimization already beat cuDNN in most cases (shown by \emph{Generic}).  Then, depending on the height and width values, \rewriter fuses the height and width dimensions to save memory bandwidth (shown by \emph{+FuseDim}). Then, we split the reduction dimension K by 64 and measure the performance (\emph{+SplitK}).  Finally, we let \rewriter to choose the sizes for these 3 optimizations and measure performance (shown by \emph{+Tune}).

We observe that SplitK leads to the maximal speedup, as it leads to significant parallelism and keeps the Tensor Cores busy. More than 70\% of the kernels can get high performance by employing fusion and parallelizing the reduction dimension. Similar to CPUs, the additional speedup by tuning is small.

UNIT cannot outperform cuDNN on \#1 and \#15, because the strided data accesses lead to less data locality. However, since these adversarial cases (both CPU and GPU) only occupy a very small portion among all these models, we can still outperform vendor-provided libraries because of the generality of our optimization.

\begin{table}[t]
\vspace{-0.1in}
\captionof{table}{\textsc{Characteristics of the selected convolution layers.}}\label{tab:conv-shape}
\vspace{0in}
\def\arraystretch{1}
\setlength{\tabcolsep}{1pt}
\scriptsize
\fontsize{8pt}{10pt}\selectfont
\begin{tabular}{l|p{0.42cm}|p{0.42cm}|p{0.47cm}|p{0.42cm}|p{0.42cm}|p{0.42cm}|p{0.42cm}|p{0.47cm}|p{0.42cm}|p{0.42cm}|p{0.42cm}|p{0.47cm}|p{0.42cm}|p{0.42cm}|p{0.42cm}|p{0.42cm}}
\toprule
        & 1   & 2   & 3    & 4   & 5   & 6   & 7   & 8    & 9   & 10  & 11  & 12   & 13  & 14 & 15  & 16  \\
\midrule
C       & 288 & 160 & {\fontsize{7pt}{10pt}\selectfont 1056} & 80  & 128 & 192 & 256 & {\fontsize{7pt}{10pt}\selectfont 1024} & 128 & 576 & 96  & {\fontsize{7pt}{10pt}\selectfont 1024} & 576 & 64 & 64  & 608 \\
IHW     & 35  & 9   & 7    & 73  & 16  & 16  & 16  & 14   & 16  & 14  & 16  & 14   & 14  & 29 & 56  & 14  \\
K       & 384 & 224 & 192  & 192 & 128 & 192 & 256 & 512  & 160 & 192 & 128 & 256  & 128 & 96 & 128 & 192 \\
R=S     & 3   & 3   & 1    & 3   & 3   & 3   & 3   & 1    & 3   & 1   & 3   & 1    & 1   & 3  & 1   & 1   \\
Stride  & 2   & 1   & 1    & 1   & 1   & 1   & 1   & 1    & 1   & 1   & 1   & 1    & 1   & 1  & 2   & 1   \\  
OHW     & 17  & 7   & 7    & 71  & 14  & 14  & 14  & 14   & 14  & 14  & 14  & 14   & 14  & 27 & 28  & 14  \\
\bottomrule
\end{tabular}
\end{table}

\subsection{Extensibility} \label{sec:perf-app}
We evaluate the extensibility of \system in two aspects: to new hardware platforms and to new deep learning tensor operations. We observe that by just representing the semantics of the new tensorized instruction in tensor DSL, \system can easily extend to new tensorized instructions and tensor operations.

\begin{figure}[t]
    \centering
    \includegraphics[width=0.98\linewidth]{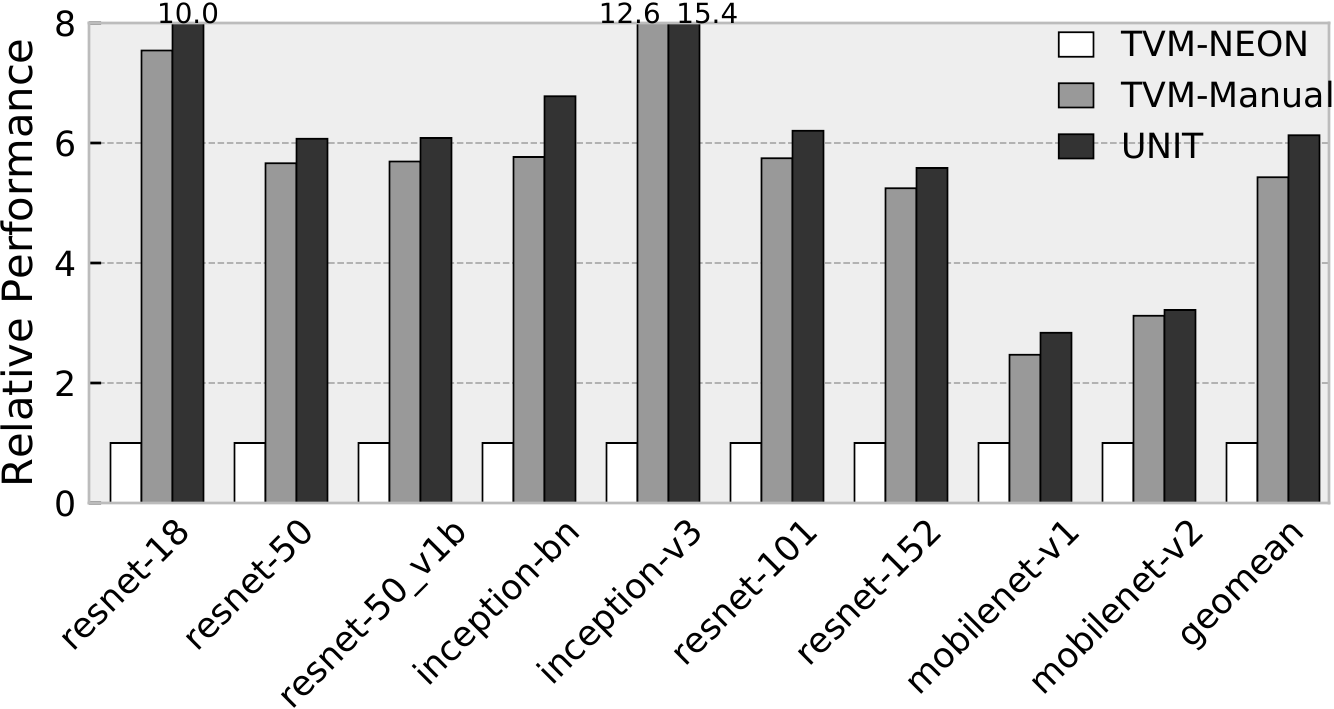}
    \vspace{-0.05in}
    \caption{The performance of ARM on model inference.}
    \label{fig:e2e-arm}
    \vspace{-0.1in}
\end{figure}

\noindent \textbf{New Hardware Platforms:} To demonstrate the capability of extending to new hardware platforms, we apply \system to an ARM CPU supporting the ARM DOT instruction. To the best of our knowledge, there is a lack of a deep learning framework with well-integrated ARM backend library support. In the absence of a framework baseline, we choose TVM compiling to ARM Neon assembly as the baseline (shown by TVM-NEON). Additionally, we find that TVM has manually-written schedules using ARM DOT instructions, which forms our second comparison baseline (shown by TVM-Manual). Note that in contrast to \system's automatic approach, this is a manually written schedule requiring intense engineering efforts. Finally, we represent the semantics of ARM DOT instruction in \system's tensor DSL and use \system to compile the models. The findings of this experiment are shown in Figure~\ref{fig:e2e-arm}, showing normalized speedup compared to the TVM-Neon baseline. The results show that \system consistently outperforms both TVM-NEON and TVM-Manual, proving \system's effectiveness in extending to new hardware platforms.

\begin{figure}[t]
    \centering
    \includegraphics[width=0.98\linewidth]{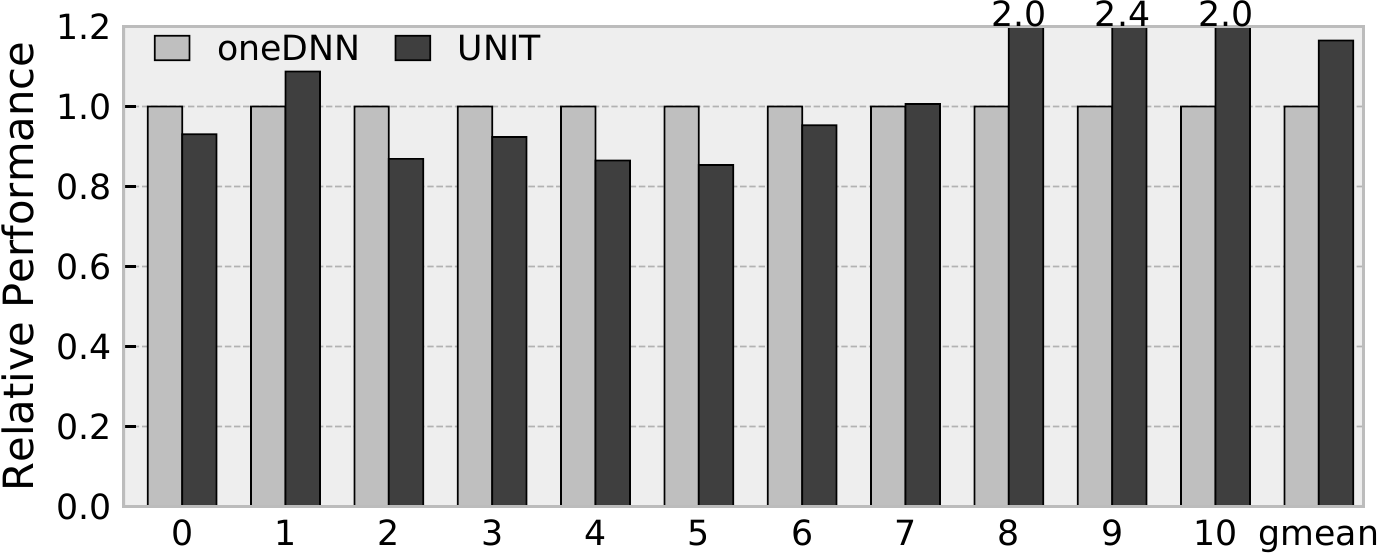}
    \vspace{-0.05in}
    \caption{The performance of each layer on res18-3d.}
    \vspace{0in}
    \label{fig:conv3d}
\end{figure}

\noindent \textbf{3D Convolution:} We test \system on 3D convolution operation for mapping Intel VNNI tensorized instructions. Note that this does not require any changes from \system perspective; we are just giving a new input (tensor-level IR for conv3d) to \system. To evaluate this extensibility, we take all the 2D convolutions from Resnet18 and manually convert them to 3D convolutions. We then apply \system on these kernels and show the speedup compared to oneDNN baseline in Figure~\ref{fig:conv3d}. We observe that \system easily extends to 3D convolution, as it has comparable performance for many convolution kernels, with an average of 1.2$\times$ speedup. 

\vspace{-0.05in}
\section{Related Work} \label{sec:rel}

\noindent\textbf{Compilation support for hardware intrinsics:}
There exists a large body of literature on compilation support for various hardware intrinsics~\cite{la-slp,poly-simd,multi-auto-vec,swizzle,simd-pldi,slp-pldi,interleave-simd,tc-cim,polydl,nv-auto-tc}. Existing production compilers such as GCC and LLVM implement auto-vectorization to leverage SIMD intrinsics. Prior works such as~\cite{poly-simd,la-slp} propose various approaches to further improve the performance of the auto-vectorizer. These approaches cannot be extended to support tensor computation intrinsics which introduce ``horizontal computation'' within each lane. 
TVM~\cite{tvm} implements an extensible interface to support new hardware intrinsics that are not limited to SIMD instructions. However, programmers need to transform the program to match the behavior of the intrinsics and declare the lowering rule for the intrinsics prior to compilation. TVM will match the computation and replace it with the code snippets that call the target hardware intrinsics. Compared to TVM, \system performs the code detection and transformation automatically. This achieves higher flexibility and productivity. There are some prior works that, similar to \system, also perform program transformation and code generation automatically for tensor computation~\cite{polydl,nv-auto-tc}. 
However, these are limited to one platform or certain intrinsics and hence are not as flexible as \system. 

\noindent \textbf{Decoupled Computation and Data Access:} The analysis pass of UNIT is inspired by the decoupled-access execute (DAE) architectures \cite{maeri,plasticine,stream-dataflow,dsagen,spu,revel,decouple-stream}. Computation and data access are decoupled and specialized separately. The computation is offloaded onto a programmable data path and data access is encoded in hardware intrinsics and executed on specialized address generation unit (AGU). \system adopts a reversed approach, it matches computation on a fixed data path, and analyzes data access fed to the data path.

\noindent \textbf{Polyhedral model:}
Many prior works have built program analysis and transformation frameworks based on the polyhedral model for tensor programs~\cite{poly-simd,polydl,nv-auto-tc,tc-cim,tensor-comprehension,polysa,mlir,ppcg}. Loop Tactics~\cite{loop-tactic} is one representative work which matches the pre-defined computation patterns in the polyhedral IR and transforms the matched patterns to optimized programs. \system distinguishes itself from Loop Tactics in: 1) Compared with the schedule tree~\cite{schedule-tree} in the polyhedral model, the tensor DSL provides more information such as loop reduction properties and operand types; 2) \system provides an end-to-end solution including auto-tuning to obtain the optimal performance, whereas Loop Tactics requires the optimized schedules to be provided manually.

\noindent\textbf{Deep learning frameworks:}
\system is complementary to the existing deep learning frameworks. Existing frameworks such as Tensorflow~\cite{tensorflow}, PyTorch~\cite{pytorch}, and MXNet~\cite{mxnet} rely on vendor-crafted libraries to support the new tensor intrinsics. TVM~\cite{tvm} requires code re-writing at the user side. \system is able to handle new operators which might not be covered by the vendor libraries and spare the user from having to perform manual re-writing. We have demonstrated the effectiveness of the methodology of \system based on TVM. Similar technique can be applied to other frameworks to further boost their performance.

\vspace{-0.1in}
\section{Conclusion} \label{sec:conc}
\vspace{-0.05in}

Deep learning has prompted hardware vendors to add specialized tensorized instructions for dense tensor operations. These instructions perform ``horizontal reduction'' accumulate elementwise computation. While promising, introducing this new idiom complicates its general purpose applicability, as one has to rely on hand-written kernels to gain high performance brought by these instructions. 
In this paper, we introduce \system, a unified compilation pipeline, that represents the tensorized instructions from different hardware platforms using the same IR, then automatically detects the applicability of the tensorized instructions in a given tensor operation, transforms the loop nest to enable easy mapping of the tensorized instruction, and finally rewrites the loop body with the tensorized instructions. \system enables automatic tensorized instruction compilation over a variety of hardware platforms like Intel/ARM CPUs and Nvidia GPUs. Our analysis shows that \system achieves 1.3$\times$ speedup over oneDNN (VNNI instruction), 1.75$\times$  over cuDNN (Tensor Core instruction),  and 1.13$\times$ over the manually written ARM intrinsics in TVM (DOT instruction).

\section*{Acknowledgements}
This work is supported by NSF grant CCF-1751400 and Mu Li's team at Amazon Web Services.


\appendix
\section{Artifact Appendix}

\subsection{Abstract}

This guide describes how to set up \emph{UNIT} compilation infrastructure and run the workloads we discussed in Section~\ref{sec:eval}. This guide provides instructions to:
\begin{itemize}
    \item Set up the experiment environment for \emph{UNIT} through Docker.
    \item Run end-to-end inference model shown in Figure~\ref{fig:e2e-cpu}, \ref{fig:e2e-gpu}, and \ref{fig:e2e-arm}.
    \item Run the experiments to demonstrate the effects of our tuning strategies shown in Figure~\ref{fig:cpu-dse}, and \ref{fig:gpu-dse}.
    \item Run the 3D-convolution results shown in Figure~\ref{fig:conv3d}.
\end{itemize}

Our experiments are conducted on Amazon EC2 \texttt{c5.12xlarge} for Intel VNNI, \texttt{p3.2xlarge} for Nvidia TensorCore, and \texttt{m6g.8xlarge} for ARM VDOT. To download and install our infrastructure, approximately 32GB of disk is required. We provide \texttt{Dockerfile} to set up the environment, and scripts to automatically run the experiments and plot the figures.

\subsection{Artifact Checklist}

\begin{itemize}
  \item {\bf Program: } As it is demonstatrated in Section~\ref{sec:method}, we use nine typical DNN models, including ResNet, ImageNet, and MobileNet.
  \item {\bf Compilation: } We need specific versions of TVM to run our experiments and baselines. They are included in the zip release.
  \item {\bf Data set: } The test data is included in our zip release.
  \item {\bf Runtime environment: } We run our artifact all on Ubuntu 18.04. For GPU, Nvidia GPU driver and additional runtime for Docker should be installed.
  \item {\bf Hardware: } We run our experiments on AWS EC2 instances --- \texttt{c5.12xlarge} for Intel VNNI, \texttt{p3.2xlarge} for Nvidia TensorCore, and \texttt{m6g.8xlarge} for ARM DOT.
  \item {\bf Execution: } We provide scripts to run our experiments discussed in Section~\ref{sec:eval}. It takes 2 hours to compile the models in Figure~\ref{fig:e2e-cpu}, half an hour to compile the models in Figure~\ref{fig:e2e-gpu}, and 1.4 hours to compile the models in Figure~\ref{fig:e2e-arm}. It takes half an hour to run the experiments in Figure~\ref{fig:cpu-dse}, and \ref{fig:gpu-dse}.
  \item {\bf Output: } Our scripts both run the experiments and plot the figures in PDF files.
  \item {\bf Experiments: } The results reported in our paper are generated by a physical machine, but in this artifact evaluation they all run on a virtual machine in Docker. Performance fluctuation may happen because of the overhead of virtualization.
\end{itemize}

\subsection{Description}

\subsubsection{How Delivered}
Download our Dockerfile, scripts, and model test data at \url{https://doi.org/10.5281/zenodo.4420522}.
\subsubsection{Hardware Dependencies}

\begin{itemize}
    \item \textbf{AVX512\_VNNI: } This is available on Intel CPUs with Cascadelake architecture. In this work, we use AWS EC2 \texttt{c5.12xlarge}. The CPU model is Intel(R) Xeon(R) Platinum 8275CL CPU @3.00GHz. The rate is \$2.04/hour, and it takes approximately one hour to set up the environment and 5 hours to run all the related experiments.
    \item \textbf{TensorCore: } This is avaiable on Nvidia GPUs with TensorCore extension. In this work, we use AWS EC2 \texttt{p3.2xlarge}. The GPU model is Tesla V100. Please install the GPU driver. The rate is \$3.06/hour, and it takes approximately 1 hour to set up the environment, and another one hour run all the related experiments.
    \item \textbf{ARM VDOT: } This is available on ARM CPU v8.2 with \texttt{dotprod} extension. In this work, we use AWS EC2 \texttt{m6g.8xlarge}. The CPU model is Amazon Graviton 2. The rate is \$1.232/hour, and it takes 1 hour to set up the environment and run the experiments. 
\end{itemize}

\subsubsection{Software Dependencies}

All our software dependences are installed automatically in Docker. Refer to \href{https://docs.docker.com/engine/install/ubuntu/}{this link} for Docker installation. When setting up the last step of the the package repository, do choose the proper tab for your CPU platform (x86 or ARM). Refer to \href{https://docs.nvidia.com/datacenter/cloud-native/container-toolkit/install-guide.html#docker}{this} to install Docker that runs Nvidia GPU. Nvidia Docker requires GPU driver installed, use this command to install:
\begin{lstlisting}
$ sudo apt-get install nvidia-driver-455
\end{lstlisting}

\subsection{Installation}
Unzip the downloaded file, and there are three sub-zips --- \texttt{tensorcore.zip}, \texttt{vnni.zip}, and \texttt{arm.zip} to evaluate the three platform we discussed in this paper.

\subsection{Experiment Workflow}

\subsubsection{GPU}
We run the TensorCore experiment on an AWS EC2 \texttt{p3.2xlarge} instance.
\begin{itemize}
    \item After building the docker image, an image hash value will be generated in the console log:
\begin{lstlisting}
$ unzip tensorcore.zip && cd tensorcore
$ sudo docker build . # 20 mins to build
$ sudo docker run -tid --runtime=nvidia <image>
$ sudo docker attach <container>
\end{lstlisting}
    \item After entering the container, the experiment scripts are all in \texttt{\$HOME} directory:
\begin{lstlisting}
$ cd $HOME
\end{lstlisting}
    \item To replicate experiments run in Figure~\ref{fig:e2e-gpu}, and \ref{fig:gpu-dse}:
\begin{lstlisting}
$ bash run_e2e.sh # Fig.9: e2e.pdf
$ bash run_ablation.sh # Fig.11: gpu-dse.pdf
\end{lstlisting}
\item It takes half an our to run these two scripts. Both the experiments and data plotting are done in these two scripts. Use these commands to take the generated PDF out of the container and look at them:
\begin{lstlisting}
$ <ctrl-p><ctrl-q> # Temporarily detach
$ sudo docker cp <container>:/root/e2e.pdf gpu-e2e.pdf
$ sudo docker cp <container>:/root/gpu-dse.pdf .
\end{lstlisting}

\end{itemize}

\subsubsection{CPU}
We run the Intel VNNI experiment on an AWS EC2 \texttt{c5.12xlarge} instance. It is also used to cross-compile ARM target.
\begin{itemize}
    \item After building the docker image, an image hash value will be generated in the console log:
\begin{lstlisting}
$ unzip vnni.zip && cd vnni
$ sudo docker build .
$ sudo docker run -tid <image>
$ sudo docker attach <container>
\end{lstlisting}
    \item After entering the container, the experiment scripts are all in \texttt{\$HOME} directory:
\begin{lstlisting}
$ cd $HOME
\end{lstlisting}
    \item To replicate experiments run in Figure~\ref{fig:e2e-cpu}, \ref{fig:cpu-dse}, and \ref{fig:conv3d}:
\begin{lstlisting}
$ bash run_e2e.sh # Fig.8: e2e.pdf
$ bash run_ablation.sh # Fig.10: cpu-dse.pdf
$ bash run_3d.sh # Fig.13: conv3d.pdf
\end{lstlisting}
\item It takes about 2.5 hours to run these experiments, and you can use the following commands to take out these plotted figures and look at them:
\begin{lstlisting}
$ <ctrl-p><ctrl-q> # Temporarily detach
$ sudo docker cp <container>:/root/e2e.pdf .
$ mv e2e.pdf cpu-e2e.pdf # Avoid conflict
$ sudo docker cp <container>:/root/gpu-dse.pdf .
$ sudo docker cp <container>:/root/conv3d.pdf .
\end{lstlisting}

\item Use the following script to run ARM target compilation:
\begin{lstlisting}
$ bash run_arm.sh
\end{lstlisting}
It takes about two hours to get all models compiled on ARM. The compiled models will be in \texttt{\$HOME/arm-base} and \texttt{\$HOME/arm-unit}.
\item Copy the compiled model to the ARM machine:
\begin{lstlisting}
$ scp -i key.pem -r arm-unit <arm-machine>:~
$ scp -i key.pem -r arm-base <arm-machine>:~
$ ssh -i key.pem <arm-machine>
\end{lstlisting}
\item Set up the ARM environment and run the experiments on ARM machine:
\begin{lstlisting}
$ unzip arm.zip && cd arm
$ mv ../arm-unit .
$ mv ../arm-base .
$ sudo docker build .
$ sudo docker run -tid <image>
$ sudo docker attach <container>
$ cd $HOME && bash run_e2e.sh
<ctrl-p> <ctrl-q>
$ sudo docker cp \
    <container>:/root/baseline.result .
$ sudo docker cp \
    <container>:/root/tensorize.result .
\end{lstlisting}
\item Bring these two \texttt{.result} files to a x86 machine, and plot the graph:
\begin{lstlisting}
$ python plot_e`2e.py baseline.result tensorize.result
# Fig. 13
$ mv e2e.pdf arm-e2e.pdf
\end{lstlisting}
\end{itemize}

\subsection{Evaluation and Expected Result}
Finally, we have these PDF files:
\begin{itemize}
    \item Figure~\ref{fig:e2e-cpu}, \ref{fig:e2e-gpu}, and \ref{fig:e2e-arm} should be compared against \texttt{cpu-e2e.pdf}, \texttt{gpu-e2e.pdf}, and \texttt{arm-e2e.pdf}.
    \begin{itemize}
        \item The ARM results reported in this paper were generated by an old version of TVM. The performance is improved in the newer version. We will fix this in camera ready.
    \end{itemize}
    \item Figure~\ref{fig:cpu-dse}, and \ref{fig:gpu-dse} should be compared against \texttt{cpu-dse.pdf}, and \texttt{gpu-dse.pdf}.
    \item Figure~\ref{fig:conv3d} should be compared against \texttt{conv3d.pdf}.
\end{itemize}

\bibliographystyle{plain}
{\footnotesize \bibliography{references}}

\end{document}